\documentclass[superscriptaddress,twocolumn]{revtex4}
\usepackage[dvips]{graphicx}

\begin{document}
\title{Gate-dependent spin-orbit coupling in multi-electron carbon nanotubes}
\author{T. S. Jespersen$^\dagger$}
\affiliation{Niels Bohr Institute \& Nano-Science Center, University
of Copenhagen, Universitetsparken 5, DK-2100 Copenhagen, Denmark}
\author{K. Grove-Rasmussen$^\dagger$}
\affiliation{Niels Bohr Institute \& Nano-Science Center, University
of Copenhagen, Universitetsparken 5, DK-2100 Copenhagen, Denmark}
\affiliation{NTT Basic Research Laboratories, NTT Corporation, 3-1
Morinosato-Wakamiya, Atsugi 243-0198, Japan}
\author{J. Paaske}
\affiliation{Niels Bohr Institute \& Nano-Science Center, University
of Copenhagen, Universitetsparken 5, DK-2100 Copenhagen, Denmark}
\author{K. Muraki}
\affiliation{NTT Basic Research Laboratories, NTT Corporation, 3-1
Morinosato-Wakamiya, Atsugi 243-0198, Japan}
\author{T. Fujisawa}
\affiliation{Research Center for Low Temperature Physics, Tokyo
Institute of Technology, Ookayama, Meguro, Tokyo 152-8551, Japan}
\author{J. Nyg{\aa}rd}
\affiliation{Niels Bohr Institute \& Nano-Science Center, University
of Copenhagen, Universitetsparken 5, DK-2100 Copenhagen, Denmark}
\author{K. Flensberg}
\affiliation{Niels Bohr Institute \& Nano-Science Center, University
of Copenhagen, Universitetsparken 5, DK-2100 Copenhagen, Denmark} 



\begin{abstract}
Understanding how the orbital motion of electrons is coupled to the
spin degree of freedom in nanoscale systems is central for
applications in spin-based electronics and quantum computation. We
demonstrate this coupling of spin and orbit in a carbon nanotube
quantum dot in the general multi-electron regime in presence of
finite disorder. Further, we find a strong systematic dependence of
the spin-orbit coupling on the electron occupation of the quantum
dot. This dependence, which even includes a sign change is not
demonstrated in any other system and follows from the
curvature-induced spin-orbit split Dirac-spectrum of the underlying
graphene lattice. Our findings unambiguously show that the spin-orbit
coupling is a general property of nanotube quantum dots which provide
a unique platform for the study of spin-orbit effects and their
applications.
\end{abstract}

\maketitle

{\bf Note: Manuscript with high resolution figures and supplement
at\\
\verb"www.fys.ku.dk/~tsand/TSJ_KGR2010.pdf"}\\ The interaction of the
spin of electrons with their orbital motion has become a focus of
attention in quantum dot research. On the one hand, this spin-orbit
interaction (SOI) provides a route for spin decoherence, which is
unwanted for purposes of quantum
computation\cite{Bulaev:2008,Fischer:2009,Churchill:2009}. On the
other hand, if properly controlled, the SOI can be utilized as a
means of electrically manipulating the spin degree of
freedom\cite{Flindt:2006,Trif:2007,Nowack:2007,Pfund:2007}.\\
\indent In this context, carbon nanotubes provide a number of
attractive features, including large confinement energies, nearly
nuclear-spin-free environment, and, most importantly, the details of
the energy level structure is theoretically well understood and
modeled, as well as experimentally highly reproducible. Remarkably,
the SOI in nanotubes was largely overlooked in the first two decades
of nanotube research and was only recently demonstrated by Kuemmeth
\emph{et al.\ }for the special case of a single carrier in
ultra-clean CNT quantum
dots\cite{Kuemmeth:2008,Kuemmeth:2010,Churchill:2009}. Except for
these reports, the SOI in nanotubes is experimentally unexplored.
Theoretically, the focus has exclusively been on the SOI-modified
band structure of disorder-free
nanotubes\cite{Ando:2000,Chico:2004,HuertasHernando:2006,Jeong:2009,Izumida:2009}.
Therefore, two important questions remain: how the effective SOI
depends on electron filling and how it appears in the general case of
quantum dots subject to
disorder. Here we answer these two questions.\\
Firstly, by low-temperature electron transport we demonstrate the
presence of a significant SOI in a disordered CNT quantum dot holding
hundreds of electrons. We identify and analyze the role of SOI in the
energy spectrum for one, two, and three electrons in the four-fold
degenerate CNT electronic shells, thus describing shells at any
electron filling. By rotating the sample, we present for the first
time spectroscopy of the same charge-states for magnetic fields both
parallel and perpendicular to the nanotube axis, thus controlling the
coupling to the orbital magnetic moment. Remarkably, a
single-electron model taking into account both SOI and disorder
quantitatively describes all essential details of the multi-electron
quantum dot spectra. Secondly, by changing the electron occupancy we
are able to tune the effective SOI and even reverse its sign in
accordance with the expected curvature-induced spin-orbit splitting
of the underlying graphene Dirac
spectrum\cite{Chico:2004,HuertasHernando:2006,Bulaev:2008,Jeong:2009,Izumida:2009}.
Such systematic dependence has not been demonstrated in any other
material system and may enable a new range of spin-orbit related
applications. This microscopic understanding and detailed modeling is
in stark contrasts to situations encountered in alternative
strong-SOI quantum-dot materials, such as InAs or InSb nanowires,
where the effective SOI arises from bulk crystal effects combined
with unknown contributions from surface effects, strain and crystal
defects\cite{Fasth:2007}. These systems often exhibit semi-random
fluctuations of, e.g.,\ the $g$-factor as single electrons are
added\cite{Csonka:2008,Nilsson:2009}. Thus, beyond fundamental
interest and the prospect of realizing recent proposals of
SOI-induced spin control in CNTs\cite{FlensbergMarcus,Bulaev:2008},
our findings pave the way for new designs of experiments utilizing
the SOI in quantum dots.\\
%
%
\section*{Zero-field splitting of the four-fold degeneracy}
Our experimental setup is presented in Fig.\ 1a. We fabricate devices
of single-wall CNTs on highly doped $\mathrm{Si}$ substrates capped
with an insulating layer of $\mathrm{SiO}_2$ (see Methods section).
The size of the quantum dots is defined by the contact separation
($400 \, \mathrm{nm}$) and the electrical properties are investigated
in a voltage biased two-terminal configuration applying a voltage
$V_{sd}$ between source-drain contacts and measuring the resulting
current $I$. The differential conductance $dI/dV_{sd}$ is measured by
standard lock-in techniques. When biased with a voltage $V_g$, the Si
substrate acts as an electrostatic gate controlling the electron
occupancy of the dot. The devices are measured at
$T=100\,\mathrm{mK}$ in a $^3\mathrm{He}/^4\mathrm{He}$ dilution
refrigerator, fitted with a piezo-rotator allowing in-plane rotations
of the device in magnetic fields
up to $9 \, \mathrm T$.\\
\indent Figure 1b shows a typical measurement of $dI/dV_{sd}$ vs.\
$V_{sd}$ and $V_g$ in the multi-electron regime of a small-band-gap
semiconducting nanotube. The pattern of diamond-shaped regions of low
conductance is expected for a quantum dot in the Coulomb blockade
regime and within each diamond the quantum dot hosts a fixed number
of electrons $N$, increasing one-by-one with increasing $V_g$. The
energy $E_{\mathrm{add}}$, required for adding a single electron
appears as the diamond heights and has been extracted in Fig.\ 1c.
The four-electron periodicity clearly observed in Fig.\ 1b and c
reflects the near four-fold degeneracy in the nanotube energy
spectrum\cite{Liang:2002,Cobden:2002}; one factor of 2 from the
intrinsic spin $(\uparrow,\downarrow)$ and one factor of 2 from the
so-called isospin $(K,K')$ that stems from the rotational symmetry of
the nanotube - electrons orbit the CNT in a clockwise or
anticlockwise direction. As is generally
observed\cite{Liang:2002,Cobden:2002,Jarillo-Herrero:2005,Makarovski:2006,Moriyama:2005},
the addition energy for the second electron in each quartet (yellow
in Fig.\ 1c) exceeds those for one and three. This was previously
interpreted as a result of disorder-induced coupling $\Delta_{KK'}$
of the clockwise and anticlockwise
states\cite{Oreg:2000,Jarillo-Herrero:2005} that splits the spectrum
into two spin-degenerate pairs of bonding/antibonding states
separated by $\Delta_{KK'}$. As mentioned, Kuemmeth \emph{et al.\
}showed recently that for the first electron in an ultra-clean
suspended nanotube quantum dot, the splitting was instead dominated
by the spin-orbit coupling. The first question we address here is
whether SOI also appears in the many-electron regime and how
it may be modified or masked by disorder.\\
\section*{Modeling spin-orbit coupling and disorder}
Performing level spectroscopy with a magnetic field $B$ applied
either parallel $(B_\|)$ or perpendicular $(B_\bot)$ to the nanotube
axis proves to be a powerful tool to analyze the separate
contributions from disorder and spin-orbit coupling. This is
illustrated in Figs.\ 2a-d, which show calculated single-particle
energy level spectra for four limiting combinations of $\Delta_{KK'}$
and the effective spin-orbit coupling $\Delta_{SO}$ (all limits are
relevant for nanotube devices depending on the degree of disorder and
CNT structure\cite{Jeong:2009,Izumida:2009}; details of the model are
provided in the Supplementary Information). In all cases a parallel
field separates the four states into pairs of increasing ($K$-like
states) and decreasing ($K'$-like states) energies. The magnitude of
the shift is given by the orbital $g$-factor $g_{\mathrm{orb}}$
reflecting the coupling of $B_\|$ to the orbital magnetic moment
caused by motion around the CNT\cite{Minot:2004}. Further, each pair
exhibits a smaller internal splitting due to the Zeeman effect.
Figure 2b shows the disorder induced coupling of $K$ and $K'$ states
resulting in an avoided crossing at $B_\|=0$ and the zero-field
splitting discussed above. In the opposite limit with SOI only (Fig.\
2c), the zero-field spectrum is also split into two doublets, but the
field dependence is markedly different and no avoided crossing
appears. In the simplest picture, this behavior originates from
coupling of the electron spin to an effective magnetic field
$\mathbf{B}_{SO} = -(\mathbf v \times \mathbf E)/c^2$ experienced by
the electron as it moves with velocity $\mathbf v$ in an electric
field $\mathbf E$. Here the speed of light, $c$, reflects the
relativistic origin of the effect. In nanotubes, the curvature of the
graphene lattice generates an effective radial electric field, and
since the velocity is mainly circumferential (and opposite for $K$
and $K'$), $\mathbf B_{SO}$ polarizes the spins along the nanotube
axis and favors parallel alignment of the spin and orbital angular
momentum. Thus, even in the absence of disorder, the spectrum splits
into two Kramers doublets $(K\downarrow,K'\uparrow)$ and
$(K\uparrow,K'\downarrow)$ separated by $\Delta_{SO}$. Interestingly,
since a perpendicular field does not couple $K$ and $K'$ the doublets
do not split along $B_\bot$ on the figure. As a consequence, the
$g$-factor, when measured in a perpendicular field, will vary from
\emph{zero} when $\Delta_{SO} \gg \Delta_{KK'}$ (Fig.\ 2c) to $2$ in
the
opposite limit (Fig.\ 2b). \\
\indent The final case, including both disorder and SOI, is of
particular importance for the present study, and the calculated
spectrum is displayed in Fig.\ 2d for $\Delta_{KK'}
> \Delta_{SO}$. Importantly, the effects of SOI are not masked
despite the dominating disorder: For parallel field, SOI remains
responsible for an asymmetric splitting of the Kramers doublets
$(\alpha,\beta)$ vs.\ $(\delta,\gamma)$, and the appearance of an
additional degeneracy in the spectrum at finite field ($\delta$ and
$\gamma$ states). In a perpendicular field, the effect of SOI is to
suppress the Zeeman splitting of the two doublets and since the
eigenstates of the SOI have spins along the nanotube axis it couples
the states with spins polarized along $B_\bot$ resulting in the
avoided crossing indicated on the figure.\\
\section*{Spin-orbit interaction revealed by spectroscopy} With this in mind we now focus
on the quartet with $4N_0 \approx 180$ electrons highlighted in Fig.\
1b and expanded in Fig.\ 3a. In order to investigate the level
structure we perform cotunneling spectroscopy, as illustrated in the
schematic Fig.\ 3b\cite{Franceschi:2001}: In Coulomb blockade,
whenever $e|V_{sd}|$ matches the energy of a transition from the
ground state $\alpha$ to an excited state $(\beta,\gamma,\delta)$,
inelastic cotunnel processes, that leave the quantum dot in the
excited state, become available for transport. This significantly
increases the current and gives rise to steps in the conductance.
These appear as gate-independent features in Fig.\ 3a (arrows) and
are clearly seen in the inset showing a trace through the center of
the one-electron $(4N_0 + 1)$ diamond along the dashed line. Thus
following the magnetic field dependence of this trace, as shown in
Fig.\ 3c, maps out the level structure. The energies of the
excitations are given by the inflection points of the curve
(\emph{i.e.\ }peaks/dips of $d^2I/dV_{sd}^2$)\cite{Paaske:2006} and
the level evolution is therefore directly evident in Fig.\ 3d-i,
which show color maps of the second derivative vs.\ $V_{sd}$ and
$B_\bot,B_\|$ for $V_g$ positioned in the center of the one, two, and
three electron charge states. As explained below, the SOI is clearly
expressed
in all three spectra.\\
\indent Consider first the one-electron case: In a parallel field
(Fig.\ 3d), the asymmetric splitting of the two doublets is evident
(black vs.\ green arrows) and applying the field perpendicularly
(Panel g), the SOI is directly expressed as the avoided crossing
indicated on the figure. The measurement is in near-perfect agreement
with the single-particle excitation spectrum calculated by
subtracting the energies of Fig.\ 3b and shown by the solid lines.
The calculation depends on only three parameters: $\Delta_{SO} = 0.15
\, \mathrm{meV}$ set directly by the avoided crossing, $\Delta_{KK'}
= 0.45 \, \mathrm{meV}$ determined from the zero-field splitting of
the doublets (see Fig.\ 2d), and $g_{\mathrm{orb}} = 5.7$ set by the
slopes of the excitation lines from $\alpha$ to $\gamma,\delta$ in
Panel d.\\
\indent Consider now the role of SOI for the doubly occupied CNT
quartet. This situation is of particular importance for quantum
computation as a paradigm for preparation of entangled states and a
fundamental part of Pauli blockade in double quantum
dots\cite{Hanson:2007}. Figures 3e,h show the measured spectra in
parallel and perpendicular fields. The model perfectly describes the
measurement and now contains no free parameters since these are fixed
by one-electron measurement. Six states are expected: the ground
state singlet-like state $\tilde S_0$ formed by the two electrons
occupying the low-energy Kramers doublet\cite{note-about-tildes},
three triplet-like $\tilde T_-,\tilde T_0,\tilde T_+$ and a
singlet-like state $\tilde S_1$, which all use one state from each
doublet and the singlet-like $\tilde S_2$ with both electrons
occupying the high-energy doublet. The ground state $\tilde S_0$ does
not appear directly in the measurement, but sets the origin for the
cotunneling excitations. The excitation to the high-energy $\tilde
S_2$ (dashed line) is absent in the experiment, since it cannot be
reached by promoting only a single electron from $\tilde S_0$. In
Fig.\ 3h excitations to $\tilde T_-$ and $\tilde T_+$ are clearly
observed, while $\tilde S_1$ and $\tilde T_0$ merge into a single
high-intensity peak showing that any exchange splitting $J$ is below
the spectroscopic line width $\approx 100 \mu
\mathrm{eV}$\cite{Moriyama:2005}. In other quartets an exchange
splitting is indeed observed (see Supplementary Information). In the
two-electron spectra the SOI is directly expressed as the avoided
crossing at $B_\bot \approx 4.5 \, \mathrm{T}$ accompanying the
$\tilde S_0 \leftrightarrow \tilde T_-$ ground state
transition\cite{note-about-S2,Fasth:2007}. In quartets of yet
stronger tunnel-coupling it is replaced by a
singlet-triplet Kondo resonance\cite{Nygard:2000} (see Supplementary Information).\\
\indent Finally, the spectrum of three electrons in the four-electron
shell is equivalent to that of a single hole in a full shell; at low
fields the $\delta$-state becomes the ground-state and $\gamma$ the
first excited state, while $\alpha$ and $\beta$ then constitute the
excited doublet. As seen by comparing Fig.\ 2b and 2d SOI breaks the
intra-shell electron-hole symmetry of the nanotube spectrum. This is
evident in the experiment when comparing Fig.\ 3d and 3f: In 3f,
increasing $B_\|$, the lowest excited state $\gamma$, barely
separates from the ground state $\delta$ and at $B_\| = 1.1 \,
\mathrm{T}$ they cross again, causing a ground state transition. At
the crossing point, the spin-degenerate ground-state results in a
zero-bias Kondo peak (see inset)\cite{Galpin:2010}. Interestingly,
this degeneracy also forms the qubit proposed in Ref.\ 1. For the
$B_\bot$-dependence the one- and three-electron cases remain
identical and Fig.\ 3i exhibits again the SOI-induced avoided crossings.\\
\section*{Gate-dependent spin-orbit coupling}
Having established the presence of SOI in the general many-electron
disordered quantum dot, we now focus on the dependence of
$\Delta_{SO}$ on the quantum dot occupation. To this end, we have
repeated the spectroscopy of Fig.\ 3 for a large number of CNT
quartets and in each case extracted $\Delta_{SO}$ by fitting to the
single-particle model (all underlying data are presented in the
Supplementary Information). Figure 4a shows the result as a function
of $V_g$. An overall decrease of $\Delta_{SO}$ is observed as
electrons are added to the conduction band and, interestingly, a
negative value is found in the valence band (\emph{i.e.} SOI favoring
anti-parallel rather than parallel spin and orbital angular momentum,
thus effectively
interchanging the one and three-electron spectra).\\
\indent The magnitude of $\Delta_{SO}$ is given by the spin-orbit
splitting of the underlying graphene band structure as we will now
discuss. For flat graphene this splitting is very weak
($\Delta_{SO}^{graphene} \sim 1 \, \mu
\mathrm{eV}$)\cite{HuertasHernando:2006} as it is second-order in the
already weak atomic SOI of carbon $\Delta_{SO}^{C} \sim 8 \,
\mathrm{meV}$. In nanotubes, however, the curvature induces a
coupling between the $\pi$- and $\sigma$-bands and generates a
curvature-induced spin-orbit splitting, which is first order in the
atomic SOI and thus greatly enhances $\Delta_{SO}$. Around a
Dirac-point of the Brillouin zone (\emph{e.g.\ }$K$) the graphene
band structure appears as on Fig.\
4b\cite{Chico:2004,HuertasHernando:2006,Jeong:2009,Izumida:2009}: The
spin-up and spin-down Dirac cones are split by SOI both in energy and
along $k_\bot$, the momentum in the circumferential direction of the
CNT. The schematic also highlights the CNT band structure (Fig.\ 4a
upper inset) obtained by imposing periodic boundary conditions on
$k_\bot$. In a finite-length CNT quantum dot also the wavevector
along the nanotube axis, $k_\|$, is quantized, and letting
$\epsilon_N$ denote the energy of the $N^\mathrm{th}$ longitudinal
mode the effective SOI for a small-gap CNT becomes
\begin{eqnarray}
\Delta_{SO,\pm} &=& E_{\uparrow}^K - E_{\downarrow}^K \\ \nonumber =
2\Delta_{SO}^0 &\pm& \sqrt{(\Delta_g+\Delta_{SO}^1)^2+\epsilon_N^2}
\\ \nonumber
&\mp& \sqrt{(\Delta_g-\Delta_{SO}^1)^2+\epsilon_N^2}.
\end{eqnarray}
Here the upper(lower) sign refers to the conduction(valence) band,
$\Delta_g$ is the curvature induced energy
gap\cite{Kane:1997,Kleiner:2001}, and the two terms $\Delta_{SO}^0$
and $\Delta_{SO}^1$ are the band structure spin-orbit parameters due
to
curvature\cite{Chico:2004,HuertasHernando:2006,Jeong:2009,Izumida:2009}.
The separate contributions of the two terms are illustrated in the
lower inset to Fig.\ 4a. $\Delta_{SO}^1$ was found already in the
work of Ando\cite{Ando:2000} and accounts for the $k_\bot$-separation
of the Dirac-cones in Fig.\ 4b. For the CNT band structure this term
acts equivalently to an Aharonov-Bohm flux from a parallel
spin-dependent magnetic field, which changes the quantization
conditions in the $k_\bot$-direction. Characteristically, its
contribution to $\Delta_{SO}$ decreases with the number of electrons
in the dot $(\epsilon_N)$ and reverses sign for the valence band.
This contrasts the $\epsilon_N$-independent contribution from the
recently predicted
$\Delta_{SO}^0$-term\cite{HuertasHernando:2006,Jeong:2009,Izumida:2009},
which acts as an effective valley-dependent Zeeman term and accounts
for the energy splitting of the Dirac-cones in Fig.\ 4b. For the
nanotube studied in Kuemmeth \emph{et al.\ }\cite{Kuemmeth:2008}
$\Delta_{SO}$ has the same sign for electrons and holes, \emph{i.e.,\
} $|\Delta_{SO}^0| > |\Delta_{SO}^1|$. In our case, the negative
values measured in the valence band demonstrate the opposite limit
$|\Delta_{SO}^0| < |\Delta_{SO}^1|$. Thus, the measured
gate-dependence of $\Delta_{SO}$ agrees with the spin-orbit splitting
of the graphene Dirac spectrum caused by the curvature of the
nanotube, and fitting to Eq.\ 1 (Fig.\ 4a, dashed line) yields
$\Delta_{SO}^0 = 10 \pm 10 \, \mu \mathrm{eV}$ and $\Delta_{SO}^1 =
220 \pm 25 \, \mu \mathrm{eV}$\cite{note-about-fitting}. Band
structure models relate these parameters to the structure of the
nanotube\cite{HuertasHernando:2006,Jeong:2009,Izumida:2009}:
$\Delta_{SO}^0 = \lambda_0 \Delta_{SO}^C \Delta_g D$ and
$\Delta_{SO}^1 = \lambda_1 \Delta_{SO}^C/D$, where $D$ is the
nanotube diameter and $\lambda_{1,2}$ constants that depend on the
CNT class (semiconducting, small-band-gap). Typically CVD-grown
single wall CNTs have diameters in the range 1-3$\, \mathrm{nm}$,
while obtaining $D$ from the measured values of
$g_{\mathrm{orb}}$\cite{note-about-diamter} gives $D\approx 6 \,
\mathrm{nm}$. Thus taking $D=1$-$6 \, \mathrm{nm}$ we estimate
$\lambda_1 = 0.03$-$0.17 \, \mathrm{nm}$ and $\lambda_0 = 0.7$-$4
\cdot 10^{-6} \, (\mathrm{nm\cdot meV})^{-1}$. While $\lambda_1$ is
consistent with the value quoted in the theoretical literature
$(0.095 \, \mathrm{nm})$\cite{Izumida:2009}, the calculated
$\lambda_0$ value $-4 \cdot 10^{-3} \, (\mathrm{nm\cdot meV})^{-1}$
does not match the experiment and similar deviations
appear\cite{Izumida:2009} when comparing the theory to the SOI values
measured in Ref.\ 8. The origin of this discrepancy remains unknown,
and further work on SOI in nanotubes with known chirality is needed
to make further progress.
\newpage

\section*{Methods}
The devices are made on a highly doped Silicon wafer terminated by
500\,nm of SiO$_2$. Alignment marks (Cr, 70\,nm) are defined by
electron beam lithography prior to deposition of catalyst islands
made of Iron nitrate (Fe(NO$_3)_3$), Molybdenum acetate and Alumina
support particles\cite{Kong:1998}. The sample is then transferred to
a furnace, where single wall carbon nanotubes are grown by chemical
vapor deposition at 850-900$^\circ$C in an atmosphere of hydrogen,
argon and methane gases. Pairs of electrodes consisting of Au/Pd
(40/10\,nm) spaced by 400\,nm are fabricated alongside the catalyst
islands by standard electron beam lithography techniques. Finally,
bonding pads (Au/Cr 150/10\,nm) are made by optical lithography and
the devices are screened by room- and
low-temperature measurements. \\
\indent We measured the sample in an Oxford dilution refrigerator
fitted with an Attocube ANRv51 piezo rotator which allows high
precision in-plane rotation of the sample in large magnetic fields.
The rotator provides resistive feedback of the actual position
measured by lock-in techniques. For electrical filtering,
room-temperature $\pi$-filters and low-temperature Thermocoax are
used. The base temperature of the modified refrigerator is around
$100\, \mathrm{mK}$. The CNT measurement setup consists of a National
Instrument digital to analog card, custom made optically coupled
amplifiers, a DL Instruments 1211 current to voltage amplifier and a
Princeton Applied Research 5210 Lock-in amplifier. Standard dc and
lock-in techniques have been used to measure current and differential
conductance $dI/dV_{sd}$ while $d^2I/dV_{sd}^2$ is obtained
numerically.\newpage

\bibliography{CNTSO}

\begin{thebibliography}{10}
\expandafter\ifx\csname url\endcsname\relax
  \def\url#1{\texttt{#1}}\fi
\expandafter\ifx\csname urlprefix\endcsname\relax\def\urlprefix{URL }\fi
\providecommand{\bibinfo}[2]{#2}
\providecommand{\eprint}[2][]{\url{#2}}

\bibitem{Bulaev:2008}
\bibinfo{author}{Bulaev, D.}, \bibinfo{author}{Trauzettel, B.} \&
  \bibinfo{author}{Loss, D.}
\newblock \bibinfo{title}{Spin-orbit interaction and anomalous spin relaxation
  in carbon nanotube quantum dots}.
\newblock \emph{\bibinfo{journal}{Phys.~Rev.~B}} \textbf{\bibinfo{volume}{77}},
  \bibinfo{pages}{235301} (\bibinfo{year}{2008}).

\bibitem{Churchill:2009}
\bibinfo{author}{Churchill, H.} \emph{et~al.}
\newblock \bibinfo{title}{Electron-nuclear interaction in c-13 nanotube double
  quantum dots}.
\newblock \emph{\bibinfo{journal}{Nat. Phys.}} \textbf{\bibinfo{volume}{5}},
  \bibinfo{pages}{321 -- 326} (\bibinfo{year}{2009}).

\bibitem{Fischer:2009}
\bibinfo{author}{Fischer, J.} \& \bibinfo{author}{Loss, D.}
\newblock \bibinfo{title}{Dealing with decoherence}.
\newblock \emph{\bibinfo{journal}{Science}} \textbf{\bibinfo{volume}{324}},
  \bibinfo{pages}{1277--1278} (\bibinfo{year}{2009}).

\bibitem{Flindt:2006}
\bibinfo{author}{Flindt, C.}, \bibinfo{author}{S{\o}rensen, A.} \&
  \bibinfo{author}{Flensberg, K.}
\newblock \bibinfo{title}{Spin-orbit mediated control of spin qubits}.
\newblock \emph{\bibinfo{journal}{Phys.~Rev.~Lett.}}
  \textbf{\bibinfo{volume}{97}}, \bibinfo{pages}{240501}
  (\bibinfo{year}{2006}).

\bibitem{Nowack:2007}
\bibinfo{author}{Nowack, K.}, \bibinfo{author}{Koppens, F.},
  \bibinfo{author}{Nazarov, Y.} \& \bibinfo{author}{Vandersypen, L.}
\newblock \bibinfo{title}{Coherent control of a single electron spin with
  electric fields}.
\newblock \emph{\bibinfo{journal}{Science}} \textbf{\bibinfo{volume}{318}},
  \bibinfo{pages}{1430 -- 1433} (\bibinfo{year}{2007}).

\bibitem{Pfund:2007}
\bibinfo{author}{Pfund, A.}, \bibinfo{author}{Shorubalko, I.},
  \bibinfo{author}{Ensslin, K.} \& \bibinfo{author}{Leturcq, R.}
\newblock \bibinfo{title}{Suppression of spin relaxation in an inas nanowire
  double quantum dot}.
\newblock \emph{\bibinfo{journal}{Phys.~Rev.~Lett.}}
  \textbf{\bibinfo{volume}{99}}, \bibinfo{pages}{036801}
  (\bibinfo{year}{2007}).

\bibitem{Trif:2007}
\bibinfo{author}{Trif, M.}, \bibinfo{author}{Golovach, V.} \&
  \bibinfo{author}{Loss, D.}
\newblock \bibinfo{title}{Spin-spin coupling in electrostatically coupled
  quantum dots}.
\newblock \emph{\bibinfo{journal}{Phys.~Rev.~B}} \textbf{\bibinfo{volume}{75}},
  \bibinfo{pages}{085307} (\bibinfo{year}{2007}).

\bibitem{Kuemmeth:2010}
\bibinfo{author}{Kuemmeth, F.}, \bibinfo{author}{Churchill, H.},
  \bibinfo{author}{Herring, P.} \& \bibinfo{author}{Marcus, C.}
\newblock \bibinfo{title}{Carbon nanotubes for coherent spintronics}.
\newblock \emph{\bibinfo{journal}{Mat. Today}} \textbf{\bibinfo{volume}{13}},
  \bibinfo{pages}{18--26} (\bibinfo{year}{2010}).

\bibitem{Kuemmeth:2008}
\bibinfo{author}{Kuemmeth, F.}, \bibinfo{author}{Ilani, S.},
  \bibinfo{author}{Ralph, D.} \& \bibinfo{author}{McEuen, P.}
\newblock \bibinfo{title}{Coupling of spin and orbital motion of electrons in
  carbon nanotubes}.
\newblock \emph{\bibinfo{journal}{Nature}} \textbf{\bibinfo{volume}{452}},
  \bibinfo{pages}{448 -- 452} (\bibinfo{year}{2008}).

\bibitem{Ando:2000}
\bibinfo{author}{Ando, T.}
\newblock \bibinfo{title}{Spin-orbit interaction in carbon nanotubes}.
\newblock \emph{\bibinfo{journal}{J. Phys. Soc. Jpn.}}
  \textbf{\bibinfo{volume}{69}}, \bibinfo{pages}{1757 -- 1763}
  (\bibinfo{year}{2000}).

\bibitem{Chico:2004}
\bibinfo{author}{Chico, L.}, \bibinfo{author}{Lopez-Sancho, M.} \&
  \bibinfo{author}{Munoz, M.}
\newblock \bibinfo{title}{Spin splitting induced by spin-orbit interaction in
  chiral nanotubes}.
\newblock \emph{\bibinfo{journal}{Phys.~Rev.~Lett.}}
  \textbf{\bibinfo{volume}{93}}, \bibinfo{pages}{176402}
  (\bibinfo{year}{2004}).

\bibitem{HuertasHernando:2006}
\bibinfo{author}{Huertas-Hernando, D.}, \bibinfo{author}{Guinea, F.} \&
  \bibinfo{author}{Brataas, A.}
\newblock \bibinfo{title}{Spin-orbit coupling in curved graphene, fullerenes,
  nanotubes, and nanotube caps}.
\newblock \emph{\bibinfo{journal}{Phys.~Rev.~B}} \textbf{\bibinfo{volume}{74}},
  \bibinfo{pages}{155426} (\bibinfo{year}{2006}).

\bibitem{Izumida:2009}
\bibinfo{author}{Izumida, W.}, \bibinfo{author}{Sato, K.} \&
  \bibinfo{author}{Saito, R.}
\newblock \bibinfo{title}{Spin-orbit interaction in single wall carbon
  nanotubes: Symmetry adapted tight-binding calculation and effective model
  analysis}.
\newblock \emph{\bibinfo{journal}{J. Phys. Soc. Jpn.}}
  \textbf{\bibinfo{volume}{78}}, \bibinfo{pages}{074707}
  (\bibinfo{year}{2009}).

\bibitem{Jeong:2009}
\bibinfo{author}{Jeong, J.} \& \bibinfo{author}{Lee, H.}
\newblock \bibinfo{title}{Curvature-enhanced spin-orbit coupling in a carbon
  nanotube}.
\newblock \emph{\bibinfo{journal}{Phys.~Rev.~B}} \textbf{\bibinfo{volume}{80}},
  \bibinfo{pages}{075409} (\bibinfo{year}{2009}).

\bibitem{Fasth:2007}
\bibinfo{author}{Fasth, C.}, \bibinfo{author}{Fuhrer, A.},
  \bibinfo{author}{Samuelson, L.}, \bibinfo{author}{Golovach, V.} \&
  \bibinfo{author}{Loss, D.}
\newblock \bibinfo{title}{Direct measurement of the spin-orbit interaction in a
  two-electron inas nanowire quantum dot}.
\newblock \emph{\bibinfo{journal}{Phys.~Rev.~Lett.}}
  \textbf{\bibinfo{volume}{98}}, \bibinfo{pages}{266801}
  (\bibinfo{year}{2007}).

\bibitem{Csonka:2008}
\bibinfo{author}{Csonka, S.} \emph{et~al.}
\newblock \bibinfo{title}{Giant fluctuations and gate control of the g-factor
  in inas nanowire quantum dots}.
\newblock \emph{\bibinfo{journal}{Nano Lett.}} \textbf{\bibinfo{volume}{8}},
  \bibinfo{pages}{3932--3935} (\bibinfo{year}{2008}).

\bibitem{Nilsson:2009}
\bibinfo{author}{Nilsson, H.} \emph{et~al.}
\newblock \bibinfo{title}{Giant, level-dependent g factors in insb nanowire
  quantum dots}.
\newblock \emph{\bibinfo{journal}{Nano Lett.}} \textbf{\bibinfo{volume}{9}},
  \bibinfo{pages}{3151--3156} (\bibinfo{year}{2009}).

\bibitem{FlensbergMarcus}
\bibinfo{author}{Flensberg, K.} \& \bibinfo{author}{Marcus, C.}
\newblock \bibinfo{title}{Bends in nanotubes allow electric spin control and
  coupling}.
\newblock \emph{\bibinfo{journal}{Phys.~Rev.~B}} \textbf{\bibinfo{volume}{81}},
  \bibinfo{pages}{195418} (\bibinfo{year}{2010}).

\bibitem{Cobden:2002}
\bibinfo{author}{Cobden, D.} \& \bibinfo{author}{Nygard, J.}
\newblock \bibinfo{title}{Shell filling in closed single-wall carbon nanotube
  quantum dots}.
\newblock \emph{\bibinfo{journal}{Phys.~Rev.~Lett.}}
  \textbf{\bibinfo{volume}{89}}, \bibinfo{pages}{046803}
  (\bibinfo{year}{2002}).

\bibitem{Liang:2002}
\bibinfo{author}{Liang, W.}, \bibinfo{author}{Bockrath, M.} \&
  \bibinfo{author}{Park, H.}
\newblock \bibinfo{title}{Shell filling and exchange coupling in metallic
  single-walled carbon nanotubes}.
\newblock \emph{\bibinfo{journal}{Phys.~Rev.~Lett.}}
  \textbf{\bibinfo{volume}{88}}, \bibinfo{pages}{126801}
  (\bibinfo{year}{2002}).

\bibitem{Jarillo-Herrero:2005}
\bibinfo{author}{Jarillo-Herrero, P.} \emph{et~al.}
\newblock \bibinfo{title}{Electronic transport spectroscopy of carbon nanotubes
  in a magnetic field}.
\newblock \emph{\bibinfo{journal}{Phys.~Rev.~Lett.}}
  \textbf{\bibinfo{volume}{94}}, \bibinfo{pages}{156802}
  (\bibinfo{year}{2005}).

\bibitem{Makarovski:2006}
\bibinfo{author}{Makarovski, A.}, \bibinfo{author}{An, L.},
  \bibinfo{author}{Liu, J.} \& \bibinfo{author}{Finkelstein, G.}
\newblock \bibinfo{title}{Persistent orbital degeneracy in carbon nanotubes}.
\newblock \emph{\bibinfo{journal}{Phys.~Rev.~B}} \textbf{\bibinfo{volume}{74}},
  \bibinfo{pages}{155431} (\bibinfo{year}{2006}).

\bibitem{Moriyama:2005}
\bibinfo{author}{Moriyama, S.}, \bibinfo{author}{Fuse, T.},
  \bibinfo{author}{Suzuki, M.}, \bibinfo{author}{Aoyagi, Y.} \&
  \bibinfo{author}{Ishibashi, K.}
\newblock \bibinfo{title}{Four-electron shell structures and an interacting
  two-electron system in carbon-nanotube quantum dots}.
\newblock \emph{\bibinfo{journal}{Phys.~Rev.~Lett.}}
  \textbf{\bibinfo{volume}{94}}, \bibinfo{pages}{186806}
  (\bibinfo{year}{2005}).

\bibitem{Oreg:2000}
\bibinfo{author}{Oreg, Y.}, \bibinfo{author}{Byczuk, K.} \&
  \bibinfo{author}{Halperin, B.}
\newblock \bibinfo{title}{Spin configurations of a carbon nanotube in a
  nonuniform externalpotential}.
\newblock \emph{\bibinfo{journal}{Phys.~Rev.~Lett.}}
  \textbf{\bibinfo{volume}{85}}, \bibinfo{pages}{365--368}
  (\bibinfo{year}{2000}).

\bibitem{Minot:2004}
\bibinfo{author}{Minot, E.}, \bibinfo{author}{Yaish, Y.},
  \bibinfo{author}{Sazonova, V.} \& \bibinfo{author}{Mceuen, P.}
\newblock \bibinfo{title}{Determination of electron orbital magnetic moments in
  carbon nanotubes}.
\newblock \emph{\bibinfo{journal}{Nature}} \textbf{\bibinfo{volume}{428}},
  \bibinfo{pages}{536 -- 539} (\bibinfo{year}{2004}).

\bibitem{Franceschi:2001}
\bibinfo{author}{De~Franceschi, S.} \emph{et~al.}
\newblock \bibinfo{title}{Electron cotunneling in a semiconductor quantum dot}.
\newblock \emph{\bibinfo{journal}{Phys.~Rev.~Lett.}}
  \textbf{\bibinfo{volume}{86}}, \bibinfo{pages}{{878--881}}
  (\bibinfo{year}{2001}).

\bibitem{Paaske:2006}
\bibinfo{author}{Paaske, J.} \emph{et~al.}
\newblock \bibinfo{title}{Non-equilibrium singlet-triplet kondo effect in
  carbon nanotubes}.
\newblock \emph{\bibinfo{journal}{Nat. Phys.}} \textbf{\bibinfo{volume}{2}},
  \bibinfo{pages}{460--464} (\bibinfo{year}{2006}).

\bibitem{Hanson:2007}
\bibinfo{author}{Hanson, R.}, \bibinfo{author}{Kouwenhoven, L.},
  \bibinfo{author}{Petta, J.}, \bibinfo{author}{Tarucha, S.} \&
  \bibinfo{author}{Vandersypen, L.}
\newblock \bibinfo{title}{Spins in few-electron quantum dots}.
\newblock \emph{\bibinfo{journal}{Rev. Mod. Phys.}}
  \textbf{\bibinfo{volume}{79}}, \bibinfo{pages}{1217--1265}
  (\bibinfo{year}{2007}).

\bibitem{note-about-tildes}
\bibinfo{note}{The states are not the conventional spin singlets and triplets
  as the they are modified by SOI as emphasized by the tildes.}

\bibitem{note-about-S2}
\bibinfo{note}{From the high-field ground state $\tilde T_-$, excitations to
  $\tilde S_2$ are actually allowed, however, the $\tilde S_0 \leftrightarrow
  \tilde T_-$ and $\tilde S_2 \leftrightarrow \tilde T_+$ avoided crossings
  occur simultaneously and as $\tilde T_-$ to $\tilde T_+$ excitations are
  forbidden the dashed line remains unseen in the experiment.}

\bibitem{Nygard:2000}
\bibinfo{author}{Nygard, J.}, \bibinfo{author}{Cobden, D.} \&
  \bibinfo{author}{Lindelof, P.}
\newblock \bibinfo{title}{Kondo physics in carbon nanotubes}.
\newblock \emph{\bibinfo{journal}{Nature}} \textbf{\bibinfo{volume}{408}},
  \bibinfo{pages}{342--346} (\bibinfo{year}{2000}).

\bibitem{Galpin:2010}
\bibinfo{author}{Galpin, M.}, \bibinfo{author}{Jayatilaka, F.},
  \bibinfo{author}{Logan, D.} \& \bibinfo{author}{Anders, F.}
\newblock \bibinfo{title}{Interplay between kondo physics and spin-orbit
  coupling in carbon nanotube quantum dots}.
\newblock \emph{\bibinfo{journal}{Phys.~Rev.~B}} \textbf{\bibinfo{volume}{81}},
  \bibinfo{pages}{075437} (\bibinfo{year}{2010}).

\bibitem{Kane:1997}
\bibinfo{author}{Kane, C.} \& \bibinfo{author}{Mele, E.}
\newblock \bibinfo{title}{Size, shape, and low energy electronic structure of
  carbon nanotubes}.
\newblock \emph{\bibinfo{journal}{Phys.~Rev.~Lett.}}
  \textbf{\bibinfo{volume}{78}}, \bibinfo{pages}{1932--1935}
  (\bibinfo{year}{1997}).

\bibitem{Kleiner:2001}
\bibinfo{author}{Kleiner, A.} \& \bibinfo{author}{Eggert, S.}
\newblock \bibinfo{title}{Band gaps of primary metallic carbon nanotubes}.
\newblock \emph{\bibinfo{journal}{Phys.~Rev.~B}} \textbf{\bibinfo{volume}{63}},
  \bibinfo{pages}{073408} (\bibinfo{year}{2001}).

\bibitem{note-about-fitting}
\bibinfo{note}{The band-gap of the device $E_g \approx 30 \, \mathrm{meV}$ is
  measured directly as a large Coulomb diamond at $V_g \approx 1 \, \mathrm{V}$
  and $\epsilon_N \approx 25 \, \mathrm{meV/V} \times V_g$ is estimated from
  the level spacing $\Delta E \approx 3 \, \mathrm{meV}$ and $\approx 8 \,
  \mathrm{shells}/V$}.

\bibitem{note-about-diamter}
\bibinfo{note}{Within the present theory the orbital $g$-factor depends on
  electron filling $g_{\mathrm{orb}} \approx (e v_F D/2\mu_B)/\sqrt{1 +
  (\epsilon_n/\Delta_g)^2}$ in agreement with the measurements (see SOM).}

\bibitem{Kong:1998}
\bibinfo{author}{Kong, J.}, \bibinfo{author}{Soh, H.},
  \bibinfo{author}{Cassell, A.}, \bibinfo{author}{Quate, C.} \&
  \bibinfo{author}{Dai, H.}
\newblock \bibinfo{title}{Synthesis of individual single-walled carbon
  nanotubes on patterned silicon wafers}.
\newblock \emph{\bibinfo{journal}{Nature}} \textbf{\bibinfo{volume}{395}},
  \bibinfo{pages}{878--881} (\bibinfo{year}{1998}).

\end{thebibliography}

\section*{Acknowledgements}
We thank P.\ E.\ Lindelof, J.\ Mygind, H.I.\ J{\o}rgensen, C.M.\
Marcus, and F.\ Kuemmeth for discussions and experimental support.
T.S.J.\ acknowledges the Carlsberg Foundation and Lundbeck Foundation
for financial support. K.G.R.\, K.F.\, J.N.\ acknowledges The Danish
Research Council and University of Copenhagen Center of Excellence.

\section*{Author contributions}
T.S.J.\ and K.G.R.\ performed the measurements, analyzed the data and
wrote the paper. T.S.J.\ designed the rotating sample stage. K.G.R.\
made the sample. K.M.,\ T.F.\ and J.N.\ participated in discussions
and writing the paper. J.P.\ and K.F.\ developed the theory and
guided the experiment.\newpage

\begin{figure*}
        \centering
        \includegraphics[width=15cm]{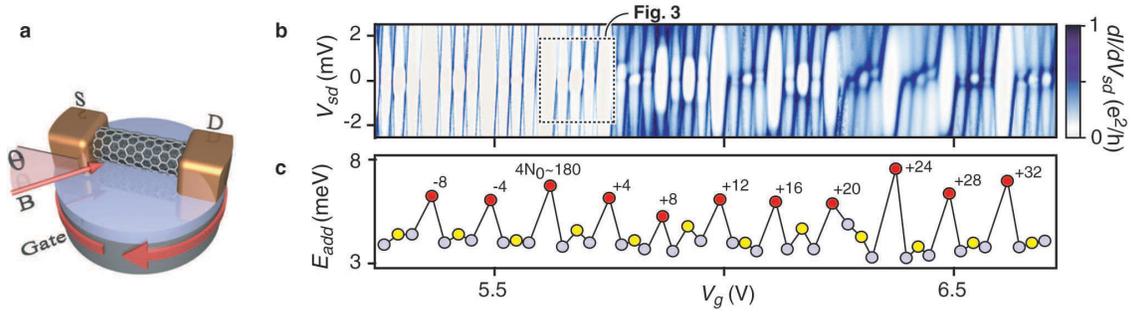}
        \caption{Four-fold periodic nanotube spectrum.
{\bf a}, Schematic illustration of the device and setup. CNT quantum
dots are measured at $T=100 \, \mathrm{mK}$ in a standard
two-terminal configuration in a cryostat modified to enable
measurements in a high magnetic field at arbitrary in-plane angles
$\theta$ to the CNT axis. {\bf b}, Typical measurement of the
differential conductance $dI/dV_{sd}$ vs.\ source-drain bias $V_{sd}$
and gate voltage $V_g$ for a multi-electron CNT quantum dot. {\bf c},
Addition energy as a function of $V_g$. In {\bf b} and {\bf c} the
characteristic filling of four-electron shells is clearly seen. }
\end{figure*}

\begin{figure*}
        \centering
        \includegraphics[width=9cm]{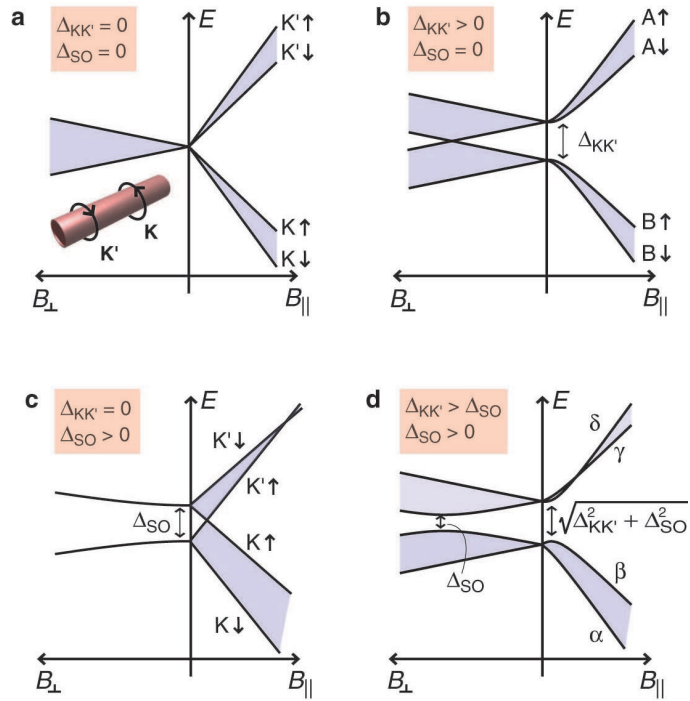}
        \caption{Role of spin-orbit interaction and disorder for the nanotube energy spectrum.
Calculated single-particle energy spectrum as a function of magnetic
field applied perpendicular $(B_{\bot})$ and parallel $(B_{\|})$ to
the CNT axis in the limiting cases of neither SOI nor disorder {\bf
a}, disorder alone {\bf b}, SOI alone {\bf c}, and the two combined
$\Delta_{KK'}
> \Delta_{SO}> 0$ {\bf d}. Depending on the CNT type, electron filling and
degree of disorder, all four situations can occur.}
\end{figure*}

\begin{figure*}
        \centering
        \includegraphics[width=16cm]{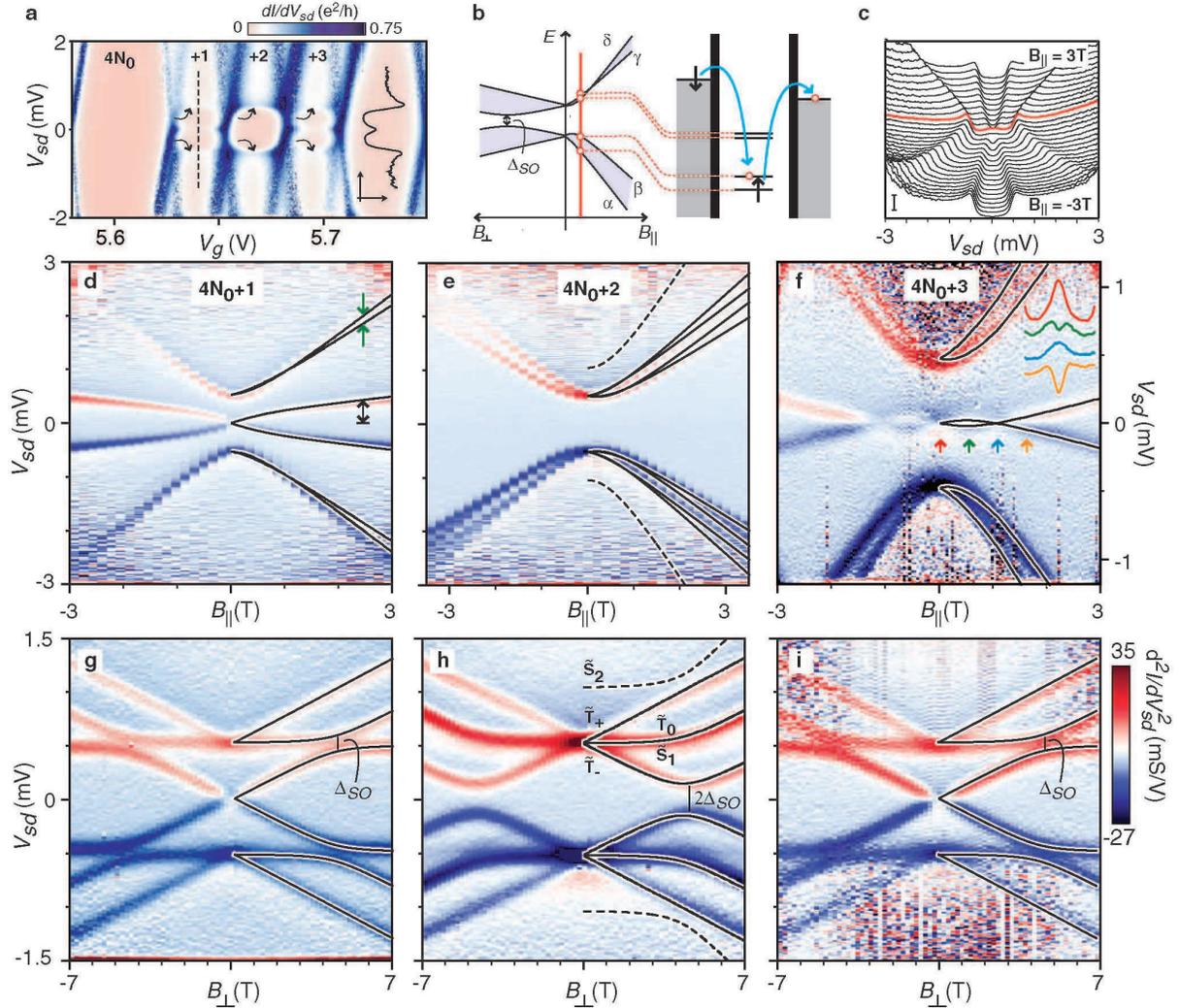}
        \caption{Spin-orbit interaction in a disordered multi-electron nanotube quantum dot.
         {\bf a}, Measurement of $dI/dV_{sd}$ vs.\ $V_{sd}$ and $V_g$
corresponding to the consecutive addition of four electrons to an
empty shell (indicated on Fig.\ 1b). A strong tunnel coupling results
in significant cotunneling which is evident as horizontal lines
truncating the diamonds (arrows). The black trace shows a cut along
the dashed line. {\bf b}, Schematic illustration of the relevant
inelastic cotunneling processes. {\bf c}, Traces along the dashed
line in {\bf a} for various $B_\|$ (red: $B=0$, scale-bar: $0.1
\mathrm{e}^2/\mathrm{h}$). {\bf d-f}, The \emph{second derivative}
$d^2I/dV_{sd}^2$ along the center of the $N_0+1,N_0+2$ and $N_0+3$
diamonds, respectively, as a function of a parallel magnetic field.
Peaks/dips appear at inflection points of the differential
conductance and thus correspond to the energy difference between
ground and excited states. In {\bf f} the inset shows $dI/dV_{sd}$
vs. $-0.3 < V_{sd} < 0.3 \, \mathrm {mV}$ and $B_\| = 0;0.55;1.1;1.65
\, \mathrm T$ (arrows) illustrating the splitting and SOI-induced
reappearance of a zero-bias Kondo resonance. {\bf g-i}, As {\bf d-f}
but measured as a function of $B_\bot$. The effective spin-orbit
coupling appears directly as the avoided crossings indicated by
$\Delta_{SO}$. In {\bf d-i} the black lines results from the
single-particle model with parameters $\Delta_{SO} = 0.15 \,
\mathrm{meV}, \Delta_{KK'} = 0.45 \, \mathrm{meV}$, and
$g_{\mathrm{orb}} = 11.4$. The dashed lines in {\bf e,h} correspond
to the excitations to the two-electron singlet-like $\tilde S_2$
state which cannot be reached by promoting a single electron from the
ground state $(\tilde S_0)$ and therefore expected to be absent in
the measurement.}
\end{figure*}

\begin{figure*}
        \centering
        \includegraphics[width=13cm]{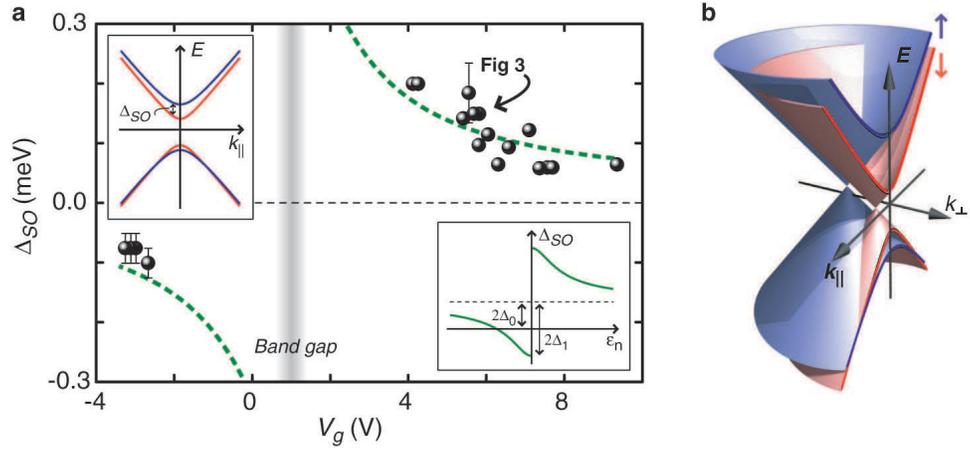}
        \caption{Tuning $\Delta_{SO}$ in accordance with the
curvature-induced spin-orbit splitting of the nanotube
Dirac-spectrum. {\bf a}, Measured effective spin-orbit coupling
strength as a function of $V_g$ extracted from spectroscopy
measurements like in Fig.\ 3, repeated for multiple shells. The
dashed line is a fit to the theory. Lower inset: Expected dependence
of $\Delta_{SO}$ on $\epsilon_N$ highlighting the two
SOI-contributions $\Delta_{SO}^0$ and $\Delta_{SO}^1$. {\bf b},
Graphene dispersion-cones around one $K$-point of the graphene
Brillouin zone. Due to SOI the spin-up (blue) and spin-down (red)
Dirac cones are split in both the vertical $(E)$ and
$k_{\bot}$-direction. The cut shows the resulting CNT band structure
also shown in the upper inset in {\bf a}.}
\end{figure*}

\end{document}